\documentclass[aps,twocolumn,amsmath,amssymb,prl,superscriptaddress,showpacs]{revtex4}
\usepackage[dvipdfmx]{graphicx}
\usepackage[dvipdfmx]{color}
\usepackage{amsmath}
\usepackage{amssymb}
\usepackage{dcolumn}
\usepackage{bm}
\usepackage{latexsym} 
\usepackage{setspace}
\usepackage{graphicx}
\usepackage{color}
\begin{document}
\title{Critical suppression of spin Seebeck effect by magnetic fields}
\author{Takashi Kikkawa}
\affiliation{Institute for Materials Research, Tohoku University, Sendai 980-8577, Japan}
\affiliation{WPI Advanced Institute for Materials Research, Tohoku University, Sendai 980-8577, Japan}
\author{Ken-ichi Uchida}
\email{kuchida@imr.tohoku.ac.jp}
\affiliation{Institute for Materials Research, Tohoku University, Sendai 980-8577, Japan}
\affiliation{PRESTO, Japan Science and Technology Agency, Saitama 332-0012, Japan}
\author{Shunsuke Daimon}
\affiliation{Institute for Materials Research, Tohoku University, Sendai 980-8577, Japan}
\author{Zhiyong Qiu}
\affiliation{WPI Advanced Institute for Materials Research, Tohoku University, Sendai 980-8577, Japan}
\affiliation{Spin Quantum Rectification Project, ERATO, Japan Science and Technology Agency, Sendai 980-8577, Japan}
\author{Yuki Shiomi}
\affiliation{Institute for Materials Research, Tohoku University, Sendai 980-8577, Japan}
\affiliation{Spin Quantum Rectification Project, ERATO, Japan Science and Technology Agency, Sendai 980-8577, Japan}
\author{Eiji Saitoh}
\affiliation{Institute for Materials Research, Tohoku University, Sendai 980-8577, Japan}
\affiliation{WPI Advanced Institute for Materials Research, Tohoku University, Sendai 980-8577, Japan}
\affiliation{Spin Quantum Rectification Project, ERATO, Japan Science and Technology Agency, Sendai 980-8577, Japan}
\affiliation{Advanced Science Research Center, Japan Atomic Energy Agency, Tokai 319-1195, Japan}
\date{\today}
\begin{abstract}
The longitudinal spin Seebeck effect (LSSE) in Pt/Y$_3$Fe$_5$O$_{12}$ (YIG) junction systems has been investigated at various magnetic fields and temperatures. We found that the LSSE voltage in a Pt/YIG-slab system is suppressed by applying high magnetic fields and this suppression is critically enhanced at low temperatures. The field-induced suppression of the LSSE in the Pt/YIG-slab system is too large at around room temperature to be explained simply by considering the effect of the Zeeman gap in magnon excitation. This result requires us to introduce a magnon-frequency-dependent mechanism into the scenario of LSSE; low-frequency magnons dominantly contribute to the LSSE. The magnetic field dependence of the LSSE voltage was observed to change by changing the thickness of YIG, suggesting that the thermo-spin conversion by the low-frequency magnons is suppressed in thin YIG films due to the long characteristic lengths of such magnons. 
\end{abstract}
\maketitle
%
%
\section{I.~~~INTRODUCTION}
%
Magnons are collective excitations of spins in magnetic ordered states, the concept of which was first introduced by Bloch in order to explain the temperature dependence of magnetization in a ferromagnet \cite{magnon}. In thermal equilibrium states, magnons behave as weakly interacting bosonic quasiparticles obeying the Bose-Einstein distribution: 
\begin{equation}\label{equ:BEfunction}
f_{\rm BE}(\epsilon, T_{\rm m})= \frac{1}{\exp(\epsilon /{k_{\rm B}}T_{\rm m})-1},
\end{equation}
where $\epsilon$ is the magnon energy, $k_{\rm B}$ is the Boltzmann constant, and $T_{\rm m}$ is the magnon temperature. In soft magnetic materials such as Y$_3$Fe$_5$O$_{12}$ (YIG) \cite{YIG_saga}, magnons are easily excited by thermal energy since the magnon dispersion is almost gapless except for a small gap due to the Zeeman effect and magnetic anisotropy ($\sim 10^{-3}$ K for YIG \cite{YIG_magnetocrystalline-anisotropy,YIG_magnetic-dipole-energy}). \par
\begin{figure*}[tb]
\begin{center}
\includegraphics{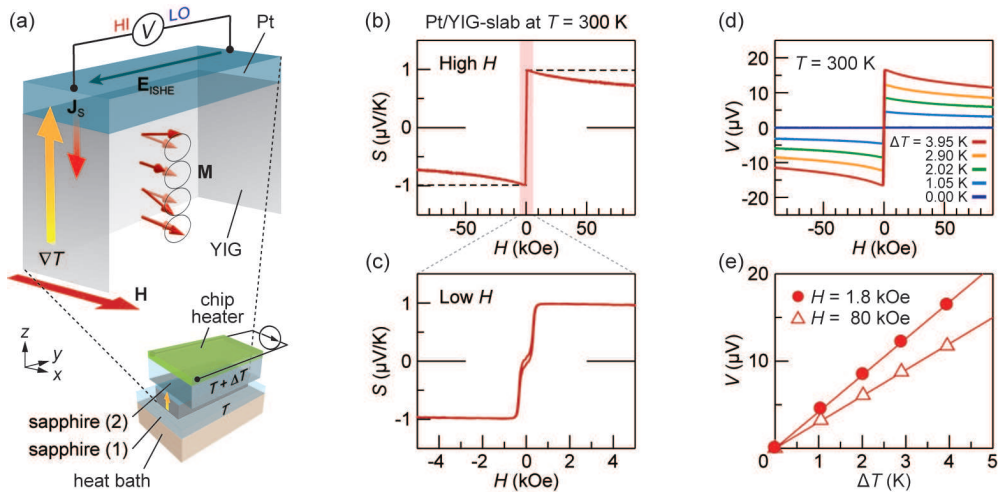}
\caption{(a) A schematic illustration of the LSSE in the Pt/YIG sample and experimental setup used in the present study. The sample is sandwiched between two sapphire plates (1) and (2). The temperatures of the sapphires (1) and (2) were respectively stabilized at $T$ and $T+\Delta T$, where $\Delta T~(>0)$ is a temperature difference. ${\bf \nabla} T$, $V$, ${\bf H}$, ${\bf M}$, ${\bf E}_{\rm ISHE}$, and ${\bf J}_{\rm s}$ denote the temperature gradient along the $+z$ direction, electric voltage between the ends of the Pt layer, magnetic field vector with the magnitude $H$, magnetization vector with the magnitude $M$, electric field induced by the ISHE, and spatial direction of the thermally generated spin current, respectively.  (b),(c) $H$ dependence of the transverse thermopower $S$ in the Pt/YIG-slab sample at $T$ = 300 K, measured when $H$ was swept between $\pm 90~\textrm{kOe}$ (b) and $\pm 5~\textrm{kOe}$ (c). (d) $H$ dependence of $V$ in the Pt/YIG-slab sample at $T$ = 300 K for various values of $\Delta T$. (e) $\Delta T$ dependence of $V$ in the Pt/YIG-slab sample at $T$ = 300 K at $H = 1.8~\textrm{kOe}$ (closed circles) and $80~\textrm{kOe}$ (open triangles). }\label{fig:1}
\end{center}
\end{figure*}
In the field of spintronics \cite{spincaloritronics1,spincaloritronics2}, magnons have attracted renewed attention, since they can carry a spin current without accompanying a charge current \cite{spincurrent,Kajiwara_nature}. Importantly, a magnon spin current in a magnet can interact with a conduction-electron spin current in an attached metal at the metal/magnet interface via the interface exchange interaction, which is described in terms of the spin-mixing conductance \cite{Tserkovnyak05,Weiler13,Qiu13}. By making use of this interaction, various spin-current-related phenomena have been developed, such as the spin pumping \cite{Kajiwara_nature,ISHE_Azevedo,ISHE_Saitoh}, spin Seebeck effect (SSE) \cite{SSE1,SSE_insulaor,SSE_Semicon1,SSE_first-LSSE,SSE_Semicon2,SSE_acousticSSE,SSE_Weiler2012PRL,SSE_2012JAPfull,SSE_Kirihara2012,SSE_Qu2013PRL,SSE_Kikkawa2013PRL,SSE_Meier2013PRB,SSE_Ramos2013APL,SSE_Uchida2013PRB,SSE_Schreier2013PRB,SSE_Kikkawa2013PRB,SSE_Schreier2013APL,SSE_Rezende2014PRB,SSE_time_resolved1,SSE_time_resolved2,SSE-JPCM,SSE_Aqeel2014JAP,SSE_Uchida2014PRX,SSE_sign,SSE_Kehlberger13}, and their reciprocal effects \cite{Kajiwara_nature,SPE1}. \par
The SSE refers to the generation of a spin current as a result of a temperature gradient in a magnetic material. Here, the thermally generated spin current is detected as electric voltage (SSE voltage) via the inverse spin Hall effect (ISHE) \cite{ISHE_Azevedo,ISHE_Saitoh,ISHE3,ISHE4,ISHE5} in a paramagnetic metal attached to a magnet. The observation of the SSE in a ferrimagnetic insulator YIG \cite{SSE_insulaor,SSE_first-LSSE} implies that this phenomenon is attributed to nonequilibrium magnon dynamics driven by a temperature gradient, since a conduction electrons' contribution in YIG is frozen out due to its large charge gap. After the pioneering theoretical work by Xiao {\it et al.} \cite{SSE_Xiao2010PRB}, the SSE is mainly described in terms of the effective magnon temperature $T_{\rm m}$ in a ferrimagnet and effective electron temperature $T_{\rm e}$ in an attached paramagnetic metal; when the effective magnon-electron temperature difference is induced by an external temperature gradient, a spin current is generated across the ferrimagnet/paramagnet interface. Adachi {\it et al.} developed linear-response theories of the magnon- and phonon-mediated SSEs \cite{SSE_Adachi2010APL,SSE_Adachi2011PRB,SSE_Adachi2013Review}. Subsequently, Hoffman {\it et al.} formulated a Landau-Lifshitz-Gilbert theory of the SSE to investigate the thickness dependence and length scale of the SSE \cite{SSE_Hoffman}. In 2014, Rezende {\it et al.} discussed the SSE in terms of a bulk magnon spin current created by a temperature gradient in a ferrimagnetic insulator \cite{SSE_Rezende2014PRB}. However, microscopic understanding of the relation between the magnon excitation and thermally generated spin currents is yet to be established, and more systematic experimental studies are necessary. \par
A clue to understand a role of magnons in SSE already manifested itself in magnetic-field-dependence measurements. In Ref. \onlinecite{SSE_Kikkawa2013PRB}, we showed that the magnitude of the SSE voltage in paramagnetic-metal (Pt, Au)/YIG-$slab$ junction systems gradually decreases with increasing the magnetic field after taking its maximum value at room temperature [see Fig. \ref{fig:1}(b)]. This suppression of the SSE voltage becomes apparent by applying high magnetic fields, while it is very small in the conventional SSE measurements in a low field range [see Fig. \ref{fig:1}(c)]. The SSE suppression by high magnetic fields is irrelevant to the anomalous Nernst effect due to static proximity ferromagnetism in Pt \cite{Proximity} since the same behavior was observed not only in Pt/YIG-slab systems but also in Au/YIG-slab systems \cite{SSE_Kikkawa2013PRB} (note that Au is free from the proximity ferromagnetism). Although this result implies that the SSE is affected by a magnon gap opening due to the Zeeman effect, there was no detailed discussion on the high-magnetic-field behavior of the SSE. In this study, using Pt/YIG systems, we systematically investigated effects of high magnetic fields on the SSE at various temperatures ranging from 300 K to 5 K. We also report the YIG-thickness dependence of the SSE voltage and its suppression at high magnetic fields. The results suggest an important role of excitation of low-frequency magnons with long characteristic lengths in the SSE, providing an important step in unraveling the nature of the SSE. \par
%
%
\section{II.~~~EXPERIMENTAL CONFIGURATION AND PROCEDURE} \label{sec:procedure}
%
%
Experiments on the SSE have been performed mainly in a longitudinal configuration owing to its simplicity \cite{SSE_first-LSSE,SSE_Weiler2012PRL,SSE_2012JAPfull,SSE_Kirihara2012,SSE_Qu2013PRL,SSE_Kikkawa2013PRL,SSE_Meier2013PRB,SSE_Ramos2013APL,SSE_Uchida2013PRB,SSE_Schreier2013PRB,SSE_Kikkawa2013PRB,SSE_Schreier2013APL,SSE_Rezende2014PRB,SSE_time_resolved1,SSE_time_resolved2,SSE-JPCM,SSE_Aqeel2014JAP,SSE_Uchida2014PRX,SSE_sign,SSE_Kehlberger13}, and we also employ this configuration in this study. Figure \ref{fig:1}(a) shows a schematic illustration of the longitudinal SSE (LSSE). In the longitudinal configuration, when a temperature gradient, $\nabla T$, is applied to a paramagnetic-metal/ferrimagnetic-insulator junction system perpendicular to the interface, a spin current is thermally generated in the paramagnetic layer along the $\nabla T$ direction. The spin current is converted into an electric field ${\bf E}_{\rm ISHE}$ by the ISHE in the paramagnetic layer if the spin-orbit interaction of the paramagnet is strong [see Fig. \ref{fig:1}(a)]. When the magnetization ${\bf M}$ of the ferrimagnet is along the $x$ direction, ${\bf E}_{\rm ISHE}$ is generated in the paramagnet along the $y$ direction following
\begin{equation}\label{equ:ISHE}
{\bf E}_{\rm ISHE} = (\theta_{\rm SH} \rho ) {\bf J}_{\rm s} \times {\bm \sigma},
\end{equation}
where $\theta_{\rm SH}$, $\rho$, ${\bf J}_{\rm s}$, and ${\bm \sigma}$ are the spin Hall angle, electric resistivity, spatial direction of a spin current, and spin-polarization vector of electrons ($||~{\bf M}$) in the paramagnet, respectively. Therefore, the LSSE can be detected electrically by measuring electric voltage $V_{\rm ISHE}$ ($=E_{\rm ISHE} L_y$) in the paramagnetic metal layer, where $E_{\rm ISHE}$ is the magnitude of ${\bf E}_{\rm ISHE}$ and $L_y$ is the length of the paramagnetic layer along the $y$ direction. \par
\begin{figure}[tb]
\begin{center}
\includegraphics{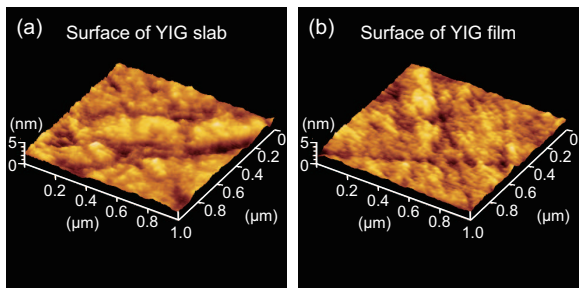}
\caption{Atomic force microscope images of the surface of the YIG-slab (a) and YIG-film (b) samples, where the surface roughness $R_{\rm a}$ is less than 0.3 nm for both the samples. All the YIG-film samples with the different thicknesses were prepared under the same growth condition by means of a liquid phase epitaxy method. After the growth, their surfaces were mechanically polished under the same condition; all the YIG films have similar surface roughness.}\label{fig:2}
\end{center}
\end{figure}
To investigate the high-magnetic-field behavior of the LSSE, we used Pt/YIG junction systems, which are now recognized as a model system for studying spin-current physics \cite{Kajiwara_nature,SSE_insulaor}. The sample used in the present study consists of a 5-nm-thick Pt film sputtered on the whole of the (111) surface of a single-crystalline YIG slab or film. The Pt films were formed on all the YIG samples at the same time.  The YIG slab has no substrate, of which the lengths along the $x$, $y$, and $z$ directions are $L_{x} = 2.0~\textrm{mm}$, $L_{y} = 4.0~\textrm{mm}$, and $L_{z} = 1.0~\textrm{mm}$, respectively. To measure the thickness dependence of the LSSE, we prepared three YIG films with the thicknesses of $t_{\rm YIG} = 10.42~\mu \textrm{m}$, $1.09~\mu \textrm{m}$, and $0.31~\mu \textrm{m}$, grown on the whole of single-crystalline Gd$_3$Ga$_5$O$_{12}$ (GGG) (111) substrates by a liquid phase epitaxy method \cite{Qiu13}. All the YIG films were prepared under the same growth condition. The GGG substrates with the YIG films were then cut into a rectangular shape with the size of $L_{x} = 2.0~\textrm{mm}$, $L_{y} = 4.0~\textrm{mm}$, and $L_{z} = 0.5~\textrm{mm}$. Before forming the Pt films, the surface of the YIG slab and films were mechanically polished with alumina powder with the diameter of 0.05$~\mu \textrm{m}$; the resultant surface roughness of the YIG slab and films are very small and comparable to each other as shown in the atomic force microscope images in Fig. \ref{fig:2}. \par 
\begin{figure}[tb]
\begin{center}
\includegraphics{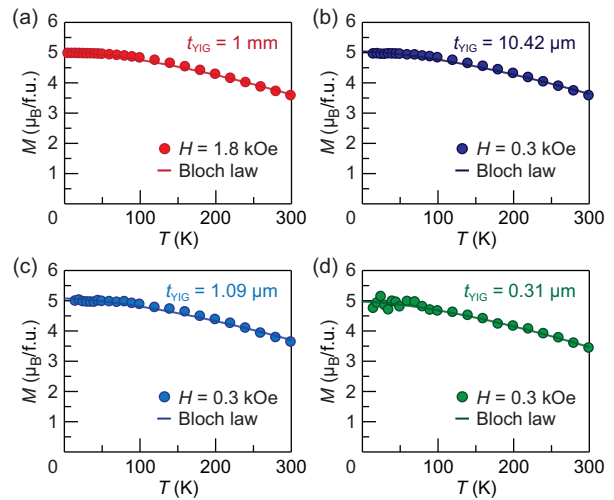}
\caption{$T$ dependence of $M$ for the YIG-slab sample with $t_{\rm YIG} = 1~\textrm{mm}$ (a) at $H = 1.8~\textrm{kOe}$ and in the YIG-film samples with $t_{\rm YIG} = 10.42~\mu \textrm{m}$ (b), $1.09~\mu \textrm{m}$ (c), and $0.31~\mu \textrm{m}$ (d) at $H = 0.3~\textrm{kOe}$, measured with a vibrating sample magnetometer. Here, $t_{\rm YIG}$ is the thickness of YIG. At $H = 1.8~\textrm{kOe}$ ($0.3~\textrm{kOe}$), the magnetization of the YIG slab (YIG films) is aligned along the $H$ direction. The $M$ values for the YIG films were extracted by subtracting the contributions from the paramagnetic GGG substrates. The $M$ data for the YIG films were detectable only for $T > 15~\textrm{K}$ because of the large paramagnetic offset coming from the GGG substrates. The solid lines were obtained by fitting the observed $M$-$T$ curves with Eq. (\ref{equ:MT-Bloch}). }\label{fig:3}
\end{center}
\end{figure}
In Fig. \ref{fig:3}, we show the temperature $T$ dependence of the magnetization $M$ for the YIG slab at $H = 1.8~\textrm{kOe}$ [Fig. \ref{fig:3}(a)] and for the YIG films at $H = 0.3~\textrm{kOe}$ [Figs. \ref{fig:3}(b)-\ref{fig:3}(d)]. Here, the $M$ values for the YIG films were obtained by subtracting the contributions from the paramagnetic GGG substrates. As $T$ decreases, the $M$ values monotonically increase and approach $\sim $ 5$~\mu_{\rm B}$, with $\mu_{\rm B}$ being the Bohr magneton, at the lowest temperature in all the YIG samples, consistent with the literature value \cite{YIG_Gilleo-Geller}. We found that the observed $T$ dependence of $M$ follows the Bloch law \cite{magnon}:
\begin{equation}\label{equ:MT-Bloch}
 M = M_0 (1-\zeta  T^{3/2}),
\end{equation}
where $M_0$ is the saturation magnetization at $T = 0~\textrm{K}$ and $\zeta $ is a constant. By fitting the experimental data with Eq. (\ref{equ:MT-Bloch}), we obtained the similar fitting parameters ($4.97~\mu_{\rm B} < M < 5.07~\mu_{\rm B}$ and $5.20 \times 10^{-5}~\textrm{K}^{-3/2} < \zeta < 5.83 \times 10^{-5}~\textrm{K}^{-3/2}$) for all the YIG samples, indicating that the magnetic property of our YIG samples is almost the same irrespective of  the YIG thickness.  \par
In the LSSE measurements, to apply $\nabla T$, the sample was sandwiched between two sapphire plates (1) and (2) [see Fig. \ref{fig:1}(a)]. The sapphire (1) is thermally connected to a heat bath of which the temperature $T$ was controlled and varied in the range from 300 K to 5 K. By applying a charge current to a chip heater attached on the top of the sapphire (2), its temperature is increased. To improve the thermal contact, thermal grease was applied between the sample and sapphire plates thinly and uniformly. The temperature difference $\Delta T$ between the sapphire (1) and (2) was measured with two thermocouples. Here, we note that the temperature gradients in the sapphire plates are negligibly small since the thermal conductivity of sapphire is much greater than that of YIG and GGG at all the temperatures \cite{Slack_Sapphire,Slack_YIG_GGG}. We also note that, since the applied $\Delta T$ is much smaller than $T$ in all the measurements [the inset to Fig. \ref{fig:7}(a)], the LSSE can be discussed within a linear-response regime [see Figs. \ref{fig:1}(d) and \ref{fig:1}(e)]. We confirmed that unintended temperature differences due to thermal artifacts are negligibly small in our measurement system in all the temperature range by checking that the LSSE disappears at $\Delta T = 0~\textrm{K}$ before each measurement. A uniform external magnetic field $H$ was applied along the $x$ direction by using a superconducting solenoid magnet, where the maximum $H$ value was 90 kOe. When $H > 1~\textrm{kOe}$ ($0.15~\textrm{kOe}$), the magnetization of the YIG slab (YIG films) is well aligned along the $H$ direction. We also confirmed that, in the range of $-90~\textrm{kOe} < H < 90~\textrm{kOe}$, the magnetoresistance ratio of the chip heater is $<0.03~\%$ in all the temperature range and the $H$ dependence of $\Delta T$ is negligibly small. Under this condition, we measured a DC electric voltage difference $V$ between the ends of the Pt layer of the Pt/YIG-slab and Pt/YIG-film samples. Hereafter, to quantitatively compare the temperature dependence of the voltage signals in different samples, we mainly plot the transverse thermopower $S \equiv (V/\Delta T)(L_z / L_{y})$. \par
%
%
\section{III.~~~RESULTS AND DISCUSSION}
\begin{figure}[tb]
\begin{center}
\includegraphics{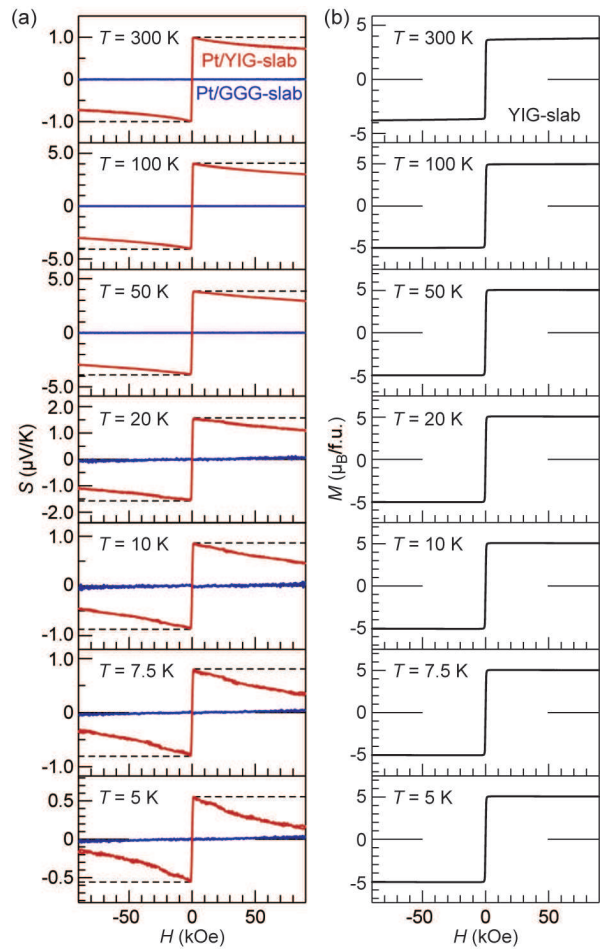}
\caption{(a) $H$ dependence of $S$ in the Pt/YIG-slab and Pt/GGG-slab samples for various values of $T$, measured when $H$ was swept between $\pm 90~\textrm{kOe}$. (b) $H$ dependence of $M$ of the YIG slab for various values of $T$, measured with a vibrating sample magnetometer. }\label{fig:4}
\end{center}
\end{figure}

%
%
Now we start by presenting the experimental results of the LSSE in the Pt/YIG-slab sample. Figure \ref{fig:4}(a) shows $S$ as a function of $H$ for various values of $T$, measured when $H$ was swept between $\pm 90~\textrm{kOe}$. When $\nabla T$ is applied to the sample, a clear $S$ signal appears due to the LSSE and its sign is reversed in response to the magnetization reversal of YIG. We found that, in the Pt/YIG-slab sample, the magnitude of the $S$ signal is suppressed by applying high magnetic fields at all the temperatures from 300 K to 5 K, while the magnitude of $M$ at each temperature is almost constant after the saturation [compare Figs. \ref{fig:4}(a) and \ref{fig:4}(b)]. This suppression cannot be explained by the normal Nernst effect \cite{Nernst1} in the Pt film since the $S$ signal in a Pt/GGG-slab sample, in which the YIG slab is replaced with a paramagnetic GGG slab, is much smaller than the $H$ dependence of the LSSE [see Fig. \ref{fig:4}(a)]. The similar $H$ dependence of the LSSE voltage was found to appear even when the thickness of the Pt layer is changed and when the Pt layer is replaced with a different metal \cite{SSE_Kikkawa2013PRB}, indicating that the magnetic-field-induced suppression of the LSSE in the Pt/YIG-slab sample is attributed to the YIG layer, not the paramagnetic metal layer. \par 
\begin{figure}[tb]
\begin{center}
\includegraphics{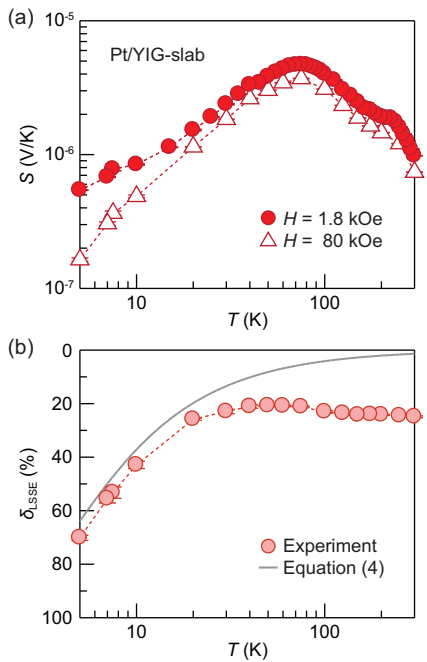}
\caption{(a) Double logarithmic plot of the $T$ dependence of $S$ in the Pt/YIG-slab sample at $H =$ 1.8 kOe (closed circles) and 80 kOe (open triangles). (b) $T$ dependence of the suppression $\delta _{\rm LSSE}$ of the LSSE voltage by magnetic fields in the Pt/YIG-slab sample (circles). Here, $\delta _{\rm LSSE} \equiv (S_{\rm max} - S_{80\textrm{kOe}})/S_{\rm max}$ with $S_{\rm max}$ and $S_{80\textrm{kOe}}$ respectively being the $S$ values at the maximum point and at $H = 80~\textrm{kOe}$. A gray line shows the $T$ dependence of $\delta _{\rm LSSE}$ calculated numerically from Eq. (\ref{equ:SSE_Calc0}) based on the conventional formulation (see Appendix). }\label{fig:5}
\end{center}
\end{figure}

In Fig. \ref{fig:5}, we show the $T$ dependence of $S$ at the positive $H$ values and of the magnetic-field-induced suppression of $S$ in the Pt/YIG-slab sample. When the sample temperature is decreased from 300 K, the magnitude of $S$ monotonically increases and reaches its maximum value around $T = 75~\textrm{K}$ [see Fig. \ref{fig:5}(a)]. On decreasing the temperature further, the $S$ signal begins to decrease and goes to zero. This $T$ dependence of the LSSE with peak structure is qualitatively consistent with previous results \cite{SSE_2012JAPfull,SSE_Rezende2014PRB}. Importantly, as shown in Fig. \ref{fig:5}(b), the suppression of the LSSE thermopower $\delta _{\rm LSSE}$ also exhibits temperature dependence in the Pt/YIG-slab sample, where $\delta _{\rm LSSE}$ is defined as $(S_{\rm max} - S_{80\textrm{kOe}})/S_{\rm max}$ with $S_{\rm max}$ and $S_{80\textrm{kOe}}$ respectively being the $S$ values at the maximum point and at $H = 80~\textrm{kOe}$. We found that the suppression of the LSSE in the Pt/YIG-slab sample is almost constant ($20~\% < \delta _{\rm LSSE} < 25~\%$) above 30 K and strongly enhanced below 30 K; the $\delta _{\rm LSSE}$ value in the Pt/YIG-slab sample reaches $\sim$70 \% at $T = 5~\textrm{K}$ [see Fig. \ref{fig:5}(b)]. \par
The critical field-induced suppression of the LSSE at low temperatures below 30 K is seemingly consistent with conventional SSE models combined with the effect of the Zeeman gap in magnon excitation. In the conventional formulation \cite{SSE_Adachi2011PRB,SSE_Adachi2013Review,SSE_Hoffman,SSE_Bender}, the LSSE voltage $V_{\rm LSSE}$ is expressed as the following factor related to the magnon excitation: 
\begin{eqnarray}\label{equ:SSE_Calc0}
V_{\rm LSSE} & \propto & \int^{\infty}_{g\mu_{\rm B}H}d\epsilon \: \epsilon \: {\mathcal D}(\epsilon, H)[f_{\rm BE}(\epsilon, T_{\rm m})-f_{\rm BE}(\epsilon, T_{\rm e})] \nonumber \\
& \propto & \int^{\infty}_{g\mu_{\rm B}H}d\epsilon \: \epsilon \: {\mathcal D}(\epsilon, H)\left.{ \frac{\partial f_{\rm BE}}{\partial T_{\rm m}}}\right|_{T_{\rm m}=T}, 
\end{eqnarray}
where ${\mathcal D(\epsilon, H)}$ is the density of states of magnons in the ferrimagnetic insulator. To obtain the differential form in Eq. (\ref{equ:SSE_Calc0}), we assume that the modulation of the effective temperatures induced by the external temperature gradient is very small ($T_{\rm m} \sim T_{\rm e}$ and $|T_{\rm m(e)} - T| \ll T$ as demonstrated in Ref. \onlinecite{SSE_Schreier2013PRB}). We numerically calculated the right-hand side of Eq. (\ref{equ:SSE_Calc0}) by assuming the density of states of parabolic exchange magnon modes: ${\mathcal D}_{\rm 0}\sqrt{\epsilon -g\mu_{\rm B}H}$ with the amplitude ${\mathcal D}_{\rm 0}$, energy $\epsilon$, $g$-factor $g$ ($= 2.0$ for YIG), and Bohr magneton $\mu_{\rm B}$, where the magnon gap due to the Zeeman effect is described as $g\mu_{\rm B}H$. This parabolic dispersion well reproduces the magnon band structure of YIG in the low-energy range ($T<30~\textrm{K}$) \cite{YIG_Plant2}. In Fig. \ref{fig:5}(b), we compare the $T$ dependence of $\delta _{\rm LSSE}$ in the Pt/YIG-slab sample with that calculated from Eq. (\ref{equ:SSE_Calc0}); below 30 K, the observed and calculated $\delta _{\rm LSSE}$ values agree with each other within the difference of 10 \% (see Appendix). \par
The inconsistency between the observed suppression of the LSSE voltage and the conventional formulation becomes apparent with increasing the temperature. Equation (\ref{equ:SSE_Calc0}) shows that the suppression of $V_{\rm LSSE}$ at $T = 300~\textrm{K}$ is smaller than 2 \% even under the high magnetic fields, which is much smaller than the experimental results as shown in Fig. \ref{fig:5}(b) ($\delta _{\rm LSSE} \sim 25~\%$ at 300 K). This is because the small Zeeman energy is defeated by thermal fluctuations when $g\mu_{\rm B}H \ll k_{\rm B} T$ (note that the magnon gap energy at $H = 80~\textrm{kOe}$ corresponds to $g\mu_{\rm B}H/k_{\rm B}= 10.7~\textrm{K} \ll 300~\textrm{K}$); to affect the magnon excitation by magnetic fields, the magnon energy has to be comparable to or less than the Zeeman energy in the conventional model. In contrast, the observed large suppression of the LSSE voltage in the Pt/YIG-slab sample indicates that the magnon excitation relevant to the LSSE is affected by magnetic fields even at around room temperature, suggesting that low-frequency magnons of which the energy is comparable to the Zeeman energy (less than $\sim 30~\textrm{K}$) provide a dominant contribution to the LSSE. \par
The importance of low-frequency magnons in the mechanism of the LSSE is clarified by focusing on their length scale. It is notable that magnons with low frequencies exhibit long thermalization (energy relaxation) lengths \cite{Sanders-Walton,Zhang-Zhang2012PRL,Zhang-Zhang2012PRB,YIG_magnon-temp,YIG_Boona-Heremans}, where magnons cannot be thermalized within the range less than the thermalization lengths. In the Pt/YIG systems under a temperature gradient, magnons can deviate from local thermal equilibrium, and the deviation becomes greater for magnons with longer thermalization lengths \cite{SSE_Xiao2010PRB,Sanders-Walton}. The frequency dependence of the magnon thermalization lengths indicates that low-frequency magnons with long thermalization lengths play a central role in the nonequilibrium states between magnons in YIG and electrons in Pt at the Pt/YIG interface. In contrast, the contribution from high-frequency magnons of which the energy is much greater than the Zeeman energy is expected to be weaker since they are closer to local thermal equilibrium due to their short thermalization lengths \cite{Zhang-Zhang2012PRL,Zhang-Zhang2012PRB,YIG_magnon-temp,YIG_Boona-Heremans}. This spectral non-uniformity of the thermo-spin conversion can be responsible for the unexpectedly strong suppression of the LSSE voltage in the Pt/YIG-slab sample, an interpretation consistent with other fragmentary pieces of information \cite{SSE_time_resolved1,SSE_time_resolved2,YIG_magnon-temp,YIG_Boona-Heremans}. Although the conventional SSE theories do not include this magnon-frequency-dependent mechanism, similar non-local nature has been introduced for phonon-electron systems as the concept of ``spectral non-uniform temperature'' in Ref. \onlinecite{SSE_Sinova}. \par
\begin{figure}[tb]
\begin{center}
\includegraphics{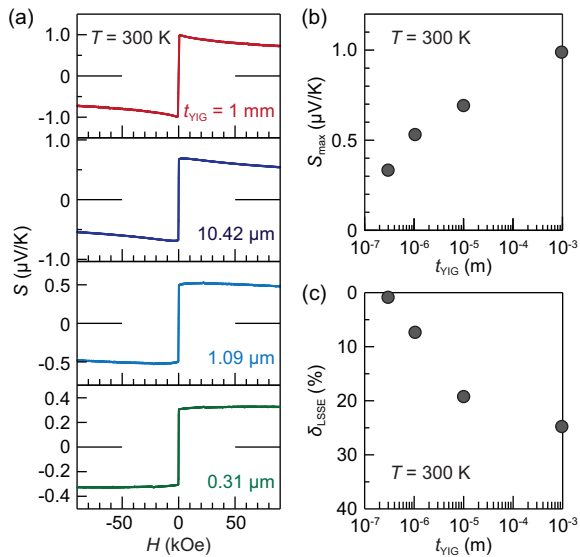}
\caption{(a) $H$ dependence of $S$ in the Pt/YIG-slab sample with the YIG thickness of $t_{\rm YIG} = 1~\textrm{mm}$ and in the Pt/YIG-film samples with $t_{\rm YIG} = 10.42~\mu \textrm{m}$, $1.09~\mu \textrm{m}$, and $0.31~\mu \textrm{m}$ at $T = 300~\textrm{K}$, measured when $H$ was swept between $\pm 90~\textrm{kOe}$. (b) $t_{\rm YIG}$ dependence of $S_{\rm max}$ at $T = 300~\textrm{K}$. (c) $t_{\rm YIG}$ dependence of $\delta _{\rm LSSE}$ at $T = 300~\textrm{K}$. }\label{fig:6}
\end{center}
\end{figure}
To verify the above scenario, we investigated the YIG-thickness dependence of the high-magnetic-field response of the LSSE. Because of the long-range nature of low-frequency magnons, the spectral non-uniformity of the thermo-spin conversion should affect the LSSE in terms of the thickness of YIG. In Fig. \ref{fig:6}(a), we compare the $H$ dependence of $S$ in the Pt/YIG-slab and Pt/YIG-film samples with different YIG thicknesses ($t_{\rm YIG} = 10.42~\mu \textrm{m}$, $1.09~\mu \textrm{m}$, and $0.31~\mu \textrm{m}$) at $T = 300~\textrm{K}$. Although we observed clear LSSE signals in all the samples, the magnitude of the LSSE thermopower monotonically decreases with decreasing $t_{\rm YIG}$. This behavior is consistent with the experimental results reported by Kirihara {\it et al.} \cite{SSE_Kirihara2012} and Kehlberger {\it et al.} \cite{SSE_Kehlberger13} [see Fig. \ref{fig:6}(b)] \cite{comment_length}. This $t_{\rm YIG}$ dependence suggests that the magnon excitation relevant to the LSSE is limited by the boundary condition in the thin YIG films. Significantly, we found that the suppression of the LSSE by high magnetic fields, $\delta _{\rm LSSE}$, also monotonically decreases with decreasing $t_{\rm YIG}$ [Fig. \ref{fig:6}(c)]; in the thinnest Pt/YIG-film sample with $t_{\rm YIG} = 0.31~\mu \textrm{m}$, the LSSE signal is almost constant for $H > 0.05~\textrm{kOe}$ ($\delta _{\rm LSSE} \sim 1~\%$ even at $H = 80~\textrm{kOe}$) [Fig. \ref{fig:6}(a)]. Similar behavior was reported in Ref. \onlinecite{SSE_Schreier2013APL}. This thickness dependence indicates that the contribution of low-frequency magnons, which govern the LSSE suppression in the Pt/YIG-slab sample, fades away in the Pt/YIG-film samples when the YIG thickness is less than their thermalization lengths \cite{comment_length2}; because the long-range magnons cannot recognize the local temperature gradient in thin YIG films, such magnons are no longer in non-equilibrium. In this condition, the LSSE suppression becomes small since only remaining high-frequency magnons with the short thermalization lengths, which have energy much greater than the Zeeman energy, contribute to the LSSE.  \par
\begin{figure*}[tb]
\begin{center}
\includegraphics{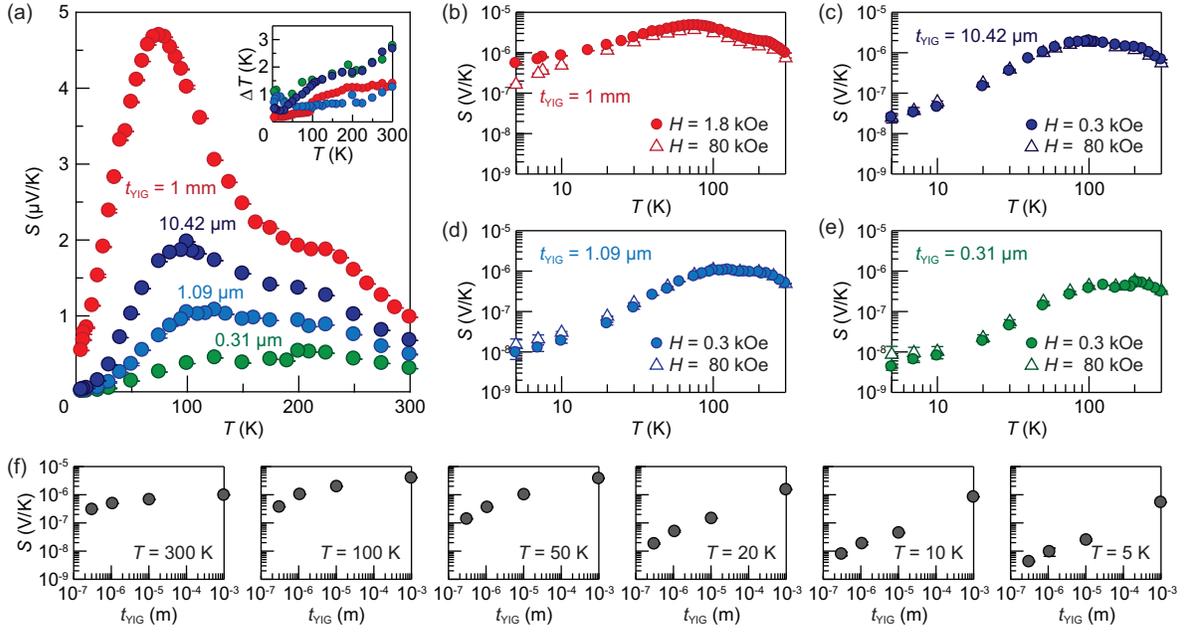}
\caption{(a) $T$ dependence of $S$ in the Pt/YIG-slab sample at $H = 1.8~\textrm{kOe}$ and in the Pt/YIG-film samples at $H = 0.3~\textrm{kOe}$. At $H = 1.8~\textrm{kOe}$ ($0.3~\textrm{kOe}$), the magnetization of the YIG slab (YIG films) is aligned along the $H$ direction, but the field-induced suppression of the LSSE is negligibly small. The inset to (a) shows the $T$ dependence of $\Delta T$ applied during the LSSE measurements. (b) Comparison between the $T$-$S$ curves at $H = 1.8~\textrm{kOe}$ (closed circles) and 80 kOe (open triangles) for the Pt/YIG-slab sample with $t_{\rm YIG} = 1~\textrm{mm}$. (c)-(e) Comparison between the $T$-$S$ curves at $H = 0.3~\textrm{kOe}$ (closed circles) and 80 kOe (open triangles) for the Pt/YIG-film samples with $t_{\rm YIG} = 10.42~\mu \textrm{m}$ (c), $1.09~\mu \textrm{m}$ (d), and $0.31~\mu \textrm{m}$ (e). (f) Double logarithmic plot of the $t_{\rm YIG}$ dependence of $S$ in the Pt/YIG-slab sample (Pt/YIG-film samples) at $H = 1.8~\textrm{kOe}$ (0.3 kOe) for various values of $T$. }\label{fig:7}
\end{center}
\end{figure*}
Finally, we show the $T$ dependence of the LSSE thermopower in the Pt/YIG-slab and Pt/YIG-film samples for various values of $t_{\rm YIG}$ \cite{DT-distribution}. As shown in Figs. \ref{fig:7}(c)-\ref{fig:7}(e), in the thin Pt/YIG-film samples, no suppression of the LSSE appears even at low temperatures that satisfy the condition $g\mu_{\rm B}H \sim k_{\rm B} T$, which is also inconsistent with the conventional formulation described by Eq. (\ref{equ:SSE_Calc0}) (note that, although $S$ in the thin Pt/YIG film samples with $t_{\rm YIG} = 0.31$ and $1.09~\mu \textrm{m}$ slightly increases with increasing $H$ at low temperatures, this behavior is attributed to the superposition of the $H$-linear component of the transverse thermoelectric voltage). The $H$ dependence of the LSSE voltage in the thin Pt/YIG-film samples can be explained by the $T$ dependence of the magnon thermalization lengths; since the thermalization length, in general, increases with decreasing $T$ \cite{Zhang-Zhang2012PRL,Zhang-Zhang2012PRB,YIG_Boona-Heremans}, the proportion of the low-frequency magnons, affected by the boundary, to the total magnon population should increase at low temperatures. This interpretation is in qualitative agreement with the $t_{\rm YIG}$ dependence of the LSSE; we found that the magnitude of $S$ monotonically decreases with reducing $t_{\rm YIG}$ in all the temperature range [Fig. \ref{fig:7}(a)] and the dependence on $t_{\rm YIG}$ of $S$ becomes stronger at lower temperatures [Fig. \ref{fig:7}(f)]. These results demonstrate again the importance of low-frequency magnons with long characteristic lengths in the mechanism of the LSSE. \par
At the end of this section, we mention a remaining task of the LSSE research. As shown in Fig. \ref{fig:7}(a), the peak position of $S$ shifts from low to high temperatures as $t_{\rm YIG}$ decreases. To understand the origin of the YIG-thickness-dependent peak shift and peak structure of the LSSE voltage, not only magnon effects \cite{YIG_Plant2,SSE_Rezende2014PRB} but also phonon effects \cite{SSE_Adachi2013Review,YIG_Boona-Heremans} should be taken into account in the mechanism of the LSSE; the separation and quantitative evaluation of these contributions are necessary. An indispensable requirement for obtaining the complete understanding of the temperature dependence of the LSSE is the precise determination of the temperature-gradient distribution in YIG-film/GGG-substrate systems, which enables the quantitative investigation of the YIG-thickness dependence of the LSSE. A challenge for the quantitative evaluation of the temperature distribution in LSSE devices is already in progress \cite{Euler2015arXiv}. \par
%
%
\section{IV.~~~CONCLUSION}
%
%
In this study, we have investigated temperature and thickness dependencies of high-magnetic-field response of the longitudinal spin Seebeck effect (LSSE) in Pt/Y$_3$Fe$_5$O$_{12}$ (YIG) junction systems. The experimental results show that the LSSE signal is suppressed by applying high magnetic fields at the temperatures ranging from 300 K to 5 K and this suppression is enhanced with decreasing the temperature in the Pt/YIG-slab system. The suppression of the LSSE appears even when the magnon gap induced by the Zeeman effect $g\mu_{\rm B}H$ is much less than the thermal energy $k_{\rm B}T$, suggesting that low-frequency magnons with energy comparable to or less than the Zeeman energy provide a dominant contribution to the LSSE rather than the higher-frequency magnons. This spectral non-uniformity of the thermo-spin conversion is associated with the characteristic lengths of the LSSE since the LSSE signal and its magnetic field dependence are strongly affected by the thickness of YIG. We anticipate that the comprehensive LSSE data reported here fill in the missing piece of the mechanism of the LSSE and lead to the development of theories of spin-current physics. \par
{\it Closing remarks:} Recently, Jin {\it et al.} \cite{SSE_Jin2015arXiv}, Ritzmann {\it et al.} \cite{SSE_Ritzmann2015arXiv}, and Guo {\it et al.} \cite{SSE_Guo2015arXiv} also reported the high-magnetic-field dependence of the LSSE in Pt/YIG systems. The experimental results and basic interpretation of these studies are consistent with those of the present study. \par
%
\section*{ACKNOWLEDGMENTS}
%
%
The authors thank S. Maekawa, H. Adachi, Y. Ohnuma, G. E. W. Bauer, J. Barker, K. Sato, J. Xiao, Y. Tserkovnyak, and S. M. Rezende for valuable discussions. This work was supported by PRESTO ``Phase Interfaces for Highly Efficient Energy Utilization,'' Japan Strategic International Cooperative Program ASPIMATT from JST, Japan, Grant-in-Aid for Scientific Research on Innovative Area ``Nano Spin Conversion Science'' (No. 26103005), Grant-in-Aid for Young Scientists (A) (No. 25707029), Grant-in-Aid for Young Scientists (B) (No. 26790038), Grant-in-Aid for Challenging Exploratory Research (No. 26600067), Grant-in-Aid for Scientific Research (A) (No. 24244051, 15H02012) from MEXT, Japan, NEC Corporation, the Tanikawa Fund Promotion of Thermal Technology, the Casio Science Promotion Foundation, and the Iwatani Naoji Foundation. T.K. is supported by JSPS through a research fellowship for young scientists (No. 15J08026). \par
%
%
\section*{APPENDIX: NUMERICAL CALCULATION OF EQUATION (\ref{equ:SSE_Calc0})}
%
%
\begin{figure}[tb]
\begin{center}
\includegraphics{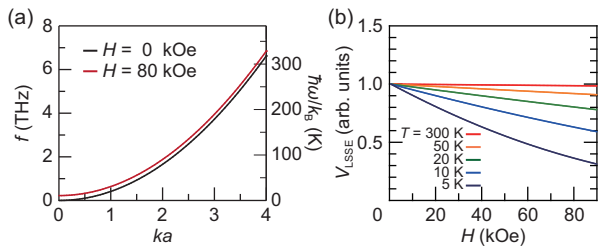}
\caption{(a) Parabolic dispersion relations of magnons at $H = 0~\textrm{kOe}$ and 80 kOe in units of the magnon frequency $f$ ($= \omega /2 \pi$) and corresponding temperature $\hbar \omega /k_{\rm B}$. The parabolic magnon dispersion includes the Zeeman gap: $\hbar \omega = Da^{2}k^{2} + g\mu_{\rm B}H$, where $\omega$ is the angular frequency, $k$ is the magnon wavenumber, $a$ is the lattice constant of YIG ($= 12.376 ~\textrm{\AA}$) \cite{YIG_Gilleo-Geller}, and $D$ is the spin-wave stiffness constant. Here, we use $Da^{2} = 4.2 \times 10^{-29} ~\textrm{erg  cm}^{2}$ \cite{YIG_Plant2,YIG_Srivastava}. (b) $H$ dependence of $V_{\rm LSSE}$ for various values of $T$, calculated from Eq. (\ref{equ:SSE_Calc0}). }\label{fig:calc}
\end{center}
\end{figure}
To clarify the $H$ dependence of the LSSE voltage $V_{\rm LSSE}$ described by the conventional formulation, we numerically calculated the right-hand side of Eq. (\ref{equ:SSE_Calc0}). For simplicity, we assume that magnons have a parabolic dispersion relation, where the density of states of magnons is affected by the Zeeman energy \cite{magnon_Kittel_text}. As shown in Fig. \ref{fig:calc}(a), the magnon gap opening due to the Zeeman effect is much smaller than thermal energy near room temperature even when the high magnetic field of $H = 80~\textrm{kOe}$ is applied. Figure \ref{fig:calc}(b) shows the calculation results of the $H$ dependence of $V_{\rm LSSE}$ for various values of $T$. We found that, although $V_{\rm LSSE}$ described by Eq. (\ref{equ:SSE_Calc0}) is suppressed by magnetic fields at low temperatures, the suppression around room temperature is much smaller than the observed $H$ dependence of the LSSE for the Pt/YIG-slab sample [compare Figs. 4(a) and \ref{fig:calc}(b)]. In Fig. \ref{fig:5}(b), we plot the $T$ dependence of $\delta _{\rm LSSE}$ calculated from Eq. (\ref{equ:SSE_Calc0}), which is defined as $(V_{\rm LSSE}^{\rm max} - V_{\rm LSSE}^{80\textrm{kOe}})/V_{\rm LSSE}^{\rm max}$ with $V_{\rm LSSE}^{\rm max}$ and $V_{\rm LSSE}^{80\textrm{kOe}}$ respectively being the $V_{\rm LSSE}$ values at the maximum point and at $H = 80~\textrm{kOe}$, indicating that the inconsistency between the experimental results and Eq. (\ref{equ:SSE_Calc0}) increases with increasing $T$. As discussed above, the small suppression of $V_{\rm LSSE}$ is attributed to the large energy difference between $g\mu_{\rm B}H$ and $k_{\rm B} T$, showing the importance of low-frequency magnons in the mechanism of the LSSE. Finally, we note that the integrand of Eq. (\ref{equ:SSE_Calc0}) consists of $\epsilon {\mathcal D} f_{\rm BE}$, while that of the magnon number is ${\mathcal D} f_{\rm BE}$ alone; since the factor $\epsilon $ in the integrand of Eq. (\ref{equ:SSE_Calc0}) eliminates the singularity of $f_{\rm BE}$ at $\epsilon = 0$, the calculated values do not reach $V_{\rm LSSE} = 0$ even when $g\mu_{\rm B}H > k_{\rm B} T$ [see Fig. \ref{fig:calc}(b)]. \par

\begin{thebibliography}{99}
\bibitem{comment_length2} Our result indicates that the characteristic length of the LSSE for low-frequency magnons, affected by the Zeeman energy in $|H| < 90~\textrm{kOe}$, is at least greater than 0.31 $\mu \textrm{m}$ at room temperature. However, to discuss the characteristic length associated with the LSSE suppression and its magnon-frequency dependence quantitatively, one has to determine the spectral non-uniformity of the thermo-spin conversion. Unfortunately, it is difficult to make this quantitative evaluation at the present stage.
\end{thebibliography}
\end{document}